\documentclass[twocolumn]{article}

\usepackage{amsmath}
\usepackage{amssymb}
\usepackage{graphicx}

\newcommand{\be}{\begin{equation}}
\newcommand{\ee}{\end{equation}}

\newcommand{\rmd}{{\mathrm d}}

\newcommand{\prob}[2]{P(#1|#2)}

\newcommand{\derivativevalue}{V}

\newcommand{\dtypestrput}{\rm put}

\newcommand{\adddtype}[2]{{#1}_{#2}}

\newcommand{\eputvalue}{\adddtype{\derivativevalue}{\dtypestrput}}

\newcommand{\strike}{E}

\begin{document}

\title{Market Implied Probability Distributions and Bayesian Skew Estimation}
\author{U. Kirchner \\ ICAP \\ PO Box 1210, Houghton, 2041, South Africa \\ulrich.kirchner@icap.co.za}

\maketitle

\begin{abstract}
We review and illustrate how the volatility smile translates into a probability distribution, the market-implied probability distribution representing believes
priced in. The effects of changes in the smile are
examined. Special attention is given to the effects of slope, which might appear at first counter-intuitive.

We then show how Bayesian methods can be used to deal with sparse real market data. With each skew in a parametric model we associate a
probability. This is illustrated with an example, for which multivariate parameter distributions are derived.
We introduce the {\em fuzzy smile} (or {\em fuzzy skew}) as a visual illustration of the skew distribution. 
\end{abstract}

\section{Introduction}

The nature of derivatives is to depend on the uncertainties associated with the underlying process. Each conceivable outcome
contributes to the current value weighted by its probability.

There are two distinct approaches to derivatives, each linked with a corresponding mathematical field.
The first one is the stochastic approach, which for example underlies the standard Black-Scholes derivation (with a special assumption about
the nature of the process). Derivative prices
are understood as expected hedging costs or profits. The process is modelled as detailed as possible, i.e., every uncertain step in time
is modelled, and the derivative value is backed out.
This is the current ``standard'' approach\footnote{The stochastic approach seems favoured as it appears to ``explain'' why derivatives
should be priced in a ``risk-neutral world''. The Black-Scholes equations can in fact be derived more easily by calculating
expected pay-offs under  a log-normal probability distribution, assuming that the mean is the risk free rate.
We think that this whole topic needs clarification and plan to address it in another paper.}, which is presented in most textbooks like \cite{wilmott}.

One important limitation of this approach is that it is only applicable to hedge-able assets in observable, liquid markets. However,
one might argue that many insurance contracts are essentially derivatives on non-observables like meteorite crash damage, and these cannot be priced
(or hedged) in a stochastic framework.

Furthermore, sometimes one knows more about the overall expected outcome than about the individual steps to get there. For example, the
central limit theorem tells us that in many cases it is reasonable to assume a Gaussian distribution for an aggregate of many
small non-Gaussian distributed random events. By modelling the individual steps one takes the risk of assuming more than one really knows.

A more general and alternative approach is to define the derivatives value as the expected pay-off. Here one does not need to model things step-by-step
through time, but rather formulate ones believes about the possible outcomes.

There are some important points to note about this. Firstly, derivatives pricing becomes now a subjective matter. People have different
believes about the future, knowledge about the current state, and needs. Hence prices should be subjective and the
fact that people trade with each other confirms this.

This is in complete contrast to the stochastic approach, which suggests that there is something like a `real' distribution describing an assets
behavior and risk. To make things worse, this `real' distribution is often estimated from historic data, ignoring obvious
(for example anthropic\footnote{The fact that one is employed in a financial market with a stock exchange means for example that
the market has not yet completely collapsed, like in Russia 1918. This however does not mean that this could not happen in the future.}) selection effects.

However, this discussion shall be left for a different time. Here we want to take a pragmatic approach and just examine which probabilities the market is
{\em pricing in}, i.e., we are concerned with what prices imply and not what we believe or what has historically happened. One could argue that the market is a big
organism of which we are just a small part. In this picture we are trying to read what this organism currently believes. We will call
these believes {\em market implied probability distributions}.

The estimation of the market implied probability distribution has been discussed by several authors.
The relationship between derivative prices and implied probabilities seems to have been first noted in \cite{breeden}.
An application to foreign currency markets was discussed in \cite{malz}, where the skew\footnote{We will use here `skew' and `smile'
interchangeably for the same thing, the observed implied volatility as a function of the strike price.}
was estimated (effectively to second order)
from traded option structures. 
There have been several other publications on specific numerical methods to deal with the noisy and sparse market data. A good overview can be found
in \cite{bahra}.

Here we want to take a different approach. First we will focus on the mathematical relationship between the skew and the implied distribution, i.e.,
we will not deal with the fitting of a skew to the data.

Then we introduce a Bayesian parameter estimation technique for the skew around at-the-money.
Here we do not just find one best-fit skew, but we consider all possible skews (numerically of course restricted to a certain large set) with an
associated probability.
As the implied probability distribution is only locally dependent on the skew we do not
need to model the far out-of-the-money skew areas (for which there is usually no data) to find the bulk
of the implied distribution around the spot level.

We present the {\em fuzzy smile} (or {\em fuzzy skew}) as an illustrative way to represent such an ``ensemble of skews''.  
Finally we combine this with the implied probability formalism, deriving a ``best guess'' based on
sparse data of what the market implied distribution could be.

 \section{Background}
The general definition of the derivatives value is the expected value of the present value of all possible cash-flows.
Here we just want to look at the simple case where discount factors are known
and cash flows occur only at one time and depend only on the final level of the underlying (no path dependence).
In this case (assuming continuous prices for convenience) 
\be
\derivativevalue = e^{-rt} \int_0^\infty p(x) f(x) \rmd x,
\ee
where $p(x) \rmd x$ is the degree of believe
(for the person
estimating the value of the derivative to himself, or priced in by the market if we have
given traded prices) that at time $t$ the underlying asset will trade in an interval $\rmd x$ around price $x$, and $f(x)$ is the pay-out in this case.

The most liquidly traded instruments are European puts and calls. For the European put above formula takes the simple form
\be
\eputvalue = e^{-rt} \int_0^\strike (\strike-x) p(x) \rmd x.
\ee
Let us re-iterate here that this is a model-independent formulation. (Models would try and tell us what $p(x)$ is.)

We can invert above functional relationship \cite{breeden}. Differentiating once gives an expression for the cumulative distribution
function
\be
\int_0^\strike p(x) \rmd x = e^{rt} \frac{\partial \eputvalue}{\partial \strike}.
\ee
Differentiating again yields the expression for the implied probability distribution
\be
p(\strike) = e^{rt} \frac{\partial^2 \eputvalue}{\partial \strike^2}.
\ee
It is important to note that this relationship is local. Hence, given the put prices
(or equivalently the volatilities) and their slope around one point one can find the
implied cumulative probability distribution at that point.

\section{An example smile}

\begin{figure}
\begin{center}
\includegraphics[width=3.2in]{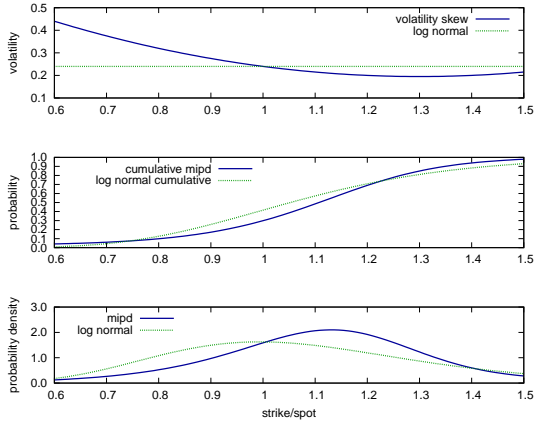}
\end{center}
\caption{We use a simple quadratic as an example skew with one year to expiry, $10\%$ interest rate, and $2\%$ dividend yield.
The green line corresponds to the log-normal distribution for the at-the-money volatility.}
\label{fig-0}
\end{figure}

We start by examining an example skew in form of a quadratic equation. The skew, the implied cumulative distribution, and the
implied probability distribution are shown in figure \ref{fig-0}, together with the log-normal distribution for the at-the-money volatility.

From the graphs it is clear that the skew causes a significant deviation from the log-normal distribution, which lies at the
heart of the Black-Scholes model.

We also note something, which might at first seem counter-intuitive:
the steep skew between $0.6$ and $1$ causes the proability density in this range to be {\em decreased}. This means, while
put-options with strikes in this range are more expensive than in a Black-Scholes world with at-the-money volatility, they are
less likely to be in-the-money on expiry!

As the put price is the (discounted) expected payoff this implies that if they are in-the-money they are more likely to be deep in the money.
``If it comes bad it comes really bad.''

\section{Effects of skew level, slope, and curvature}
Let us turn here to a more systematic study of the effects of the skew.
The implied probability density depends on the first two derivatives of the derivative prices, and hence of the skew.
The simplest model to capture all these effects and to give a reasonable description of the skew around spot is a quadratic equation, i.e.,
\be
\sigma = a + b(x-1) + c(x-1)^2,
\ee
where $x$ is the ratio of the strike over the spot price.
Here the parameters $a$, $b$, and $c$ can be interpreted as the at-the money volatility, slope, and (twice of the) curvature.

To translate the volatilities to actual put option prices we need to fix the remaining Black-scholes parameters.
Here we will use $1$ year to maturity, a dividend yield of $2\%$, and an interest rate of $10\%$.

\subsection{Effects of a level shift}

\begin{figure}
\begin{center}
\includegraphics[width=3.2in]{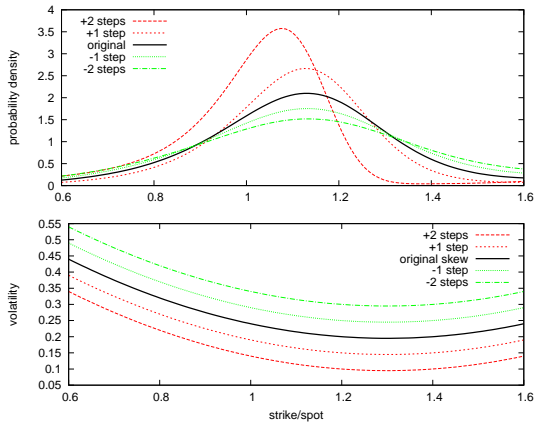}
\end{center}
\caption{Illustration of the effect of a skew level shift on the implied probability distribution.}
\label{fig-1}
\end{figure}

Figure \ref{fig-1} illustrates how a level shift of the skew affects the implied probability distribution.
While the skew causes a deviation from the log-normal distribution, the level shift still corresponds to a change in width of the
implied distribution.

\subsection{Effect of slope changes}

\begin{figure}
\begin{center}
\includegraphics[width=3.2in]{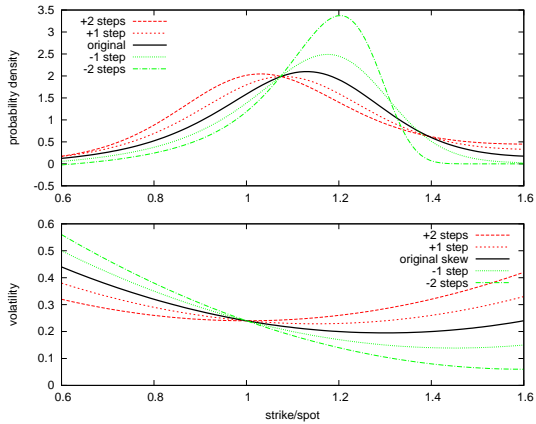}
\end{center}
\caption{Illustration of the effect of steepness on the implied probability distribution.}
\label{fig-2}
\end{figure}
The effects of slope changes on the implied probability distribution are illustrated in figure \ref{fig-2}.

It might appear at first counter-inuitive that a higher slope leads to a decreased probability density.
It is, of course, true that the higher slope (more negative) makes out-of-the money put options more expensive.
Yet, it appears that the probability of the market dropping to and below\footnote{This is not shown in figure \ref{fig-2}, but the
implied cumulative distribution is also decreased.} these levels is decreased.

\begin{figure}
\begin{center}
\includegraphics[width=3.2in]{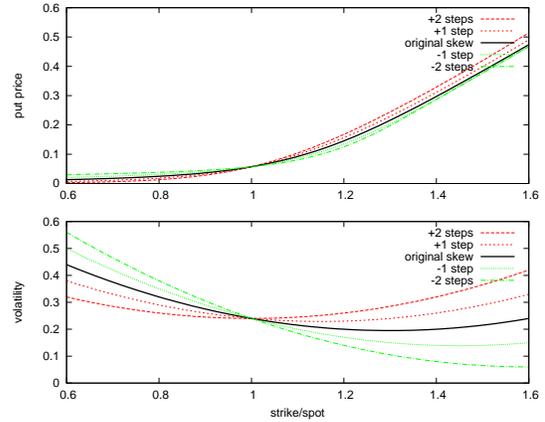}
\end{center}
\caption{A steeper slope leads to put prices which decrease less with decreasing strike. This makes narrow put-spreads cheaper.}
\label{fig-3}
\end{figure}

To understand this effect it is usefull to look at the actual put prices as shown in figure \ref{fig-3}.
Here one can see that while the higher slope leads to a levelling out of the put prices. Hence, while puts are more expensive, put spreads
are actually cheaper.
Put-spreads, however, price in the probability of the market going below the spread.

We should note here that very steep skews can break plausibility requirements.
It is clear that put prices must decrease with strike (keeping all other parameters constant).
Mathematically it is, however, possible that volatilities increase so much with decreasing strike, as to imply
rising put prices.
This just means that not every skew is a sensible model.

\subsection{Effect of curvature changes}

\begin{figure}
\begin{center}
\includegraphics[width=3.2in]{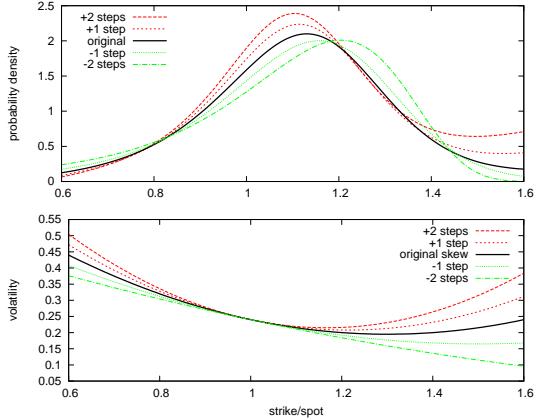}
\end{center}
\caption{Increased at-the-money curvature causes a focusing of the distribution.}
\label{fig-4}
\end{figure}

The effects of curvature changes are illustrated in figure \ref{fig-4}. We see that an increasing curvature causes a focusing of the distribution around $x=1$.

\section{Skew estimation}

In the above we assumed a given skew, here modelled by a quadratic. In real life all we have are some market quotes
and the skew itself needs to be estimated. With this estimation is associated an additional uncertainty.

Let us first define our parametric skew model as
\be
\sigma = a + b(x-1) + c(x-1)^2,
\ee
where $a$ is the at-the-money volatility, $b$ the at-the-money slope, and $c$ the (constant) curvature.

As this model is linear in $a,b,$ and $c$ a least-square best-fit can easily be found with a regression.

Baye's law tells us that
\be
\prob{M_{a,b,c}}{D I} \propto \prob{D}{M_{a,b,c}I} \prob{M_{a,b,c}}{I},
\label{equ-bayes}
\ee
where $D$ represents the observed data, $M_{a,b,c}$ the skew with parameters $a,b,c$, and $I$ any other background information.
The last term represents our {\it a-priori} beliefs about the skew parameters. Let us be conservative and assume that all
possible values are equally likely.

The first term $\prob{D}{M_{a,b,c}I}$ incorporates the information gained by observing the data $D$.
We model this by marginalising over a scale parameter and assuming a multivariate Gaussian distribution for each value
of the scale parameter\cite{jaynes}.

Above method can be used for other smile parametrisations. However, as these are usually not linear in the parameters
a regression cannot be used to find the least square best-fit.

\begin{figure}
\begin{center}
\includegraphics[width=2.8in]{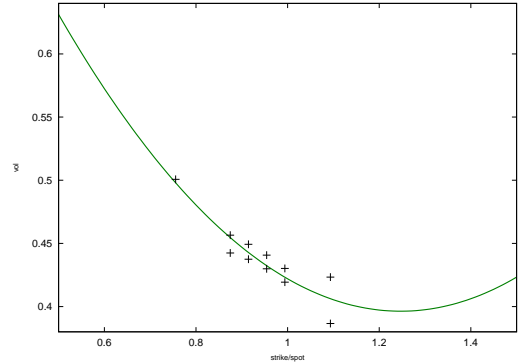}
\end{center}
\caption{Sample data with best-fit quadratic skew.}
\label{fig-10}
\end{figure}
\begin{figure}
\begin{center}
\includegraphics[width=2.8in]{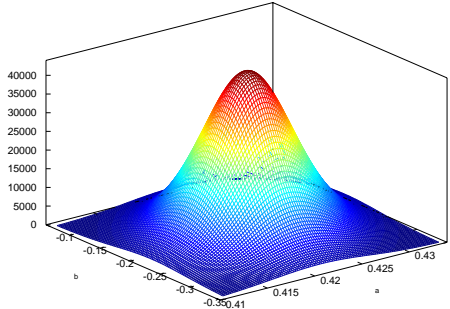}
\end{center}
\caption{Probability distribution for the skew parameters $a$ (at-the-money volatility) and $b$ (at-the-money slope). (Here parameter $c$ is integrated out.)}
\label{fig-15}
\end{figure}
\begin{figure}
\begin{center}
\includegraphics[width=2.8in]{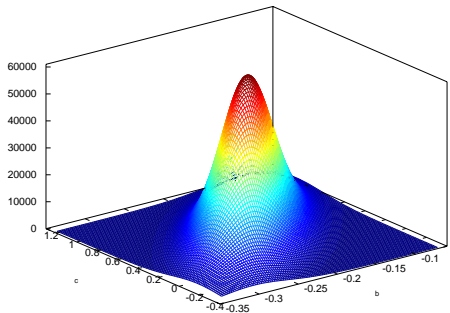}
\end{center}
\caption{Probability distribution for the skew parameters $b$ and $c$. (Here parameter $a$ is integrated out.)}
\label{fig-16}
\end{figure}

Figure \ref{fig-10} shows some example data together with a best-fit quadratic skew. We then estimate the multivariate
parameter distribution according to equation (\ref{equ-bayes}). Figures \ref{fig-15} and \ref{fig-16} show the distributions for
two ($a,b$ and $b,c$) parameters at time (the third one has been integrated out).

\begin{figure}
\begin{center}
\includegraphics[width=2.8in]{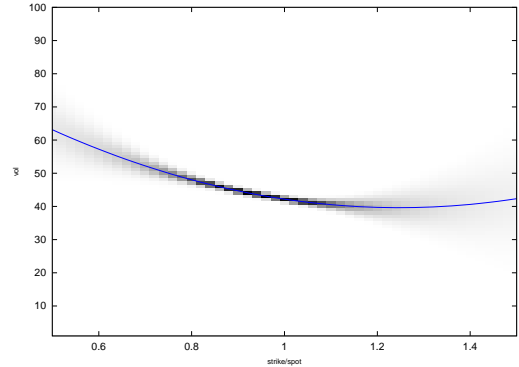}
\end{center}
\caption{Illustration of the uncertainty in the skew. The darker an area the more likely it is for the skew to go through it.
Each vertical cross section corresponds to a probability distribution for that strike.}
\label{fig-17}
\end{figure}

Hence our proposal is not to deal with a single best-fit skew/smile, but rather with a whole distribution of possible skews/smiles.
One way to illustrate this distribution of skews/smiles is the {\em fuzzy smile} as shown in figure \ref{fig-17}. Here every
vertical cut corresponds to a probability distribution with darker colors corresponding to higher probabilities. 

\begin{figure}
\begin{center}
\includegraphics[width=2.8in]{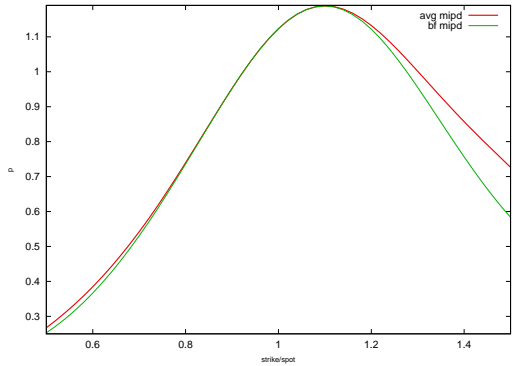}
\end{center}
\caption{Market implied proability distribution for the best-fit (green line)
and our best guess of it taking the skew fitting uncertainty into account (red line).}
\label{fig-18}
\end{figure}

Finally one can estimate the best-guess market implied probability distribution by adding up the
weighted market-implied distributions for all the skews in the ensemble. The results for our example case
are presented in figure \ref{fig-18}.

Let us note here that the skew model used is more important in the wings of the distribution. A quadratic will be
able to capture the shape around $x=1$, but will fail in the outer wings were it causes artefacts.
However, above analysis can be repeated in very much the same way for any parametric skew model. The only advantage
of a polynomial model is that the best-fit parameters can be found with regression techniques.

\section{Conclusion}

Using the Black-Scholes formula with a volatility skew is just a market-convention on how to quote option prices.
However, looking through the glasses of the Black-Scholes model into a non-Black-Scholes world can be rather confusing.
We examined and illustrated the relationship between the skew and the probability distribution it represents.

The most interesting effect is caused by the slope of the skew. The more negative the slope around a point the less the (priced in) probability to go
below the point. However, at the same time lower puts become more expensive due to the faster increasing volatility.
This implies that the probability of really large draw-downs must contribute more to the put price. 

With given sparse market data is associated an uncertainty in the skew. We showed how a Bayesian framework can be
used to take such uncertainties into account by using multi-variate parameter distributions.
A-priori beliefs about the parameters are taken into account and our estimations are hence of a subjective nature.
One can however, as was done here in the example, make very conservative assumptions for the a-priori distribution,
like an equal-likelihood distribution.

The {\em fuzzy smile/skew} was introduced as an illustrative way to visualize this ensemble of skews. It exposes in which strike
ranges we are relatively certain about the implied volatility and where we are not.

We believe that the Bayesian skew estimation, combined with the implied probability frame-work is a useful tool when dealing with
sparse option data. Above methods are easily extended to use other skew models, or incorporate more sophisticated a-priori distributions.


\begin{thebibliography}{9}

\newcommand{\bookref}[5]{#1 (#2), {\it #3}, #4,#5}
\newcommand{\articelref}[6]{#1 (#2), `#3', \textit{#4} \textbf{#5}, #6}
\newcommand{\articelrefnp}[5]{#1 (#2), `#3', \textit{#4} \textbf{#5}}

\bibitem{breeden}
\articelref{Breeden D T, Litzenberger R H}{1978}{Prices of state-contingent claims implicit in option prices}{Journal of Business}{51(4)}{621-651}

\bibitem{malz}
\articelrefnp{Malz A M}{1997}{Option-Implied Probability Distributions and Currency Excess Returns}{FRB of New York Staff Report} {32}{}

\bibitem{bahra}
\articelrefnp{Bahra B}{1997}{Implied Risk-neutral Probability Density Functions From Option Prices: Theory and Application}{Bank of England Working Paper}{66}\\
{\small (Online at {\tt http://ssrn.com/abstract=77429})}

\bibitem{jackwerth}
\articelref{Jackwerth J C}{1999}{Option Impluied Risk-Neutral Distributions and Implied Binomial Trees: A Literature Review}{Journal of Derivatives}{7(2)}{66-82}
{\small (Online at {\tt http://mpra.ub.uni-muenchen.de/11634})}

\bibitem{wilmott}
\bookref{Wilmott P}{1998}{Derivatives}{John Wiley \& Sons}{Chinchester, New York, Weinheim, Brisbane, Singapore, Toronto}


\bibitem{jaynes}
\bookref{Jaynes E T, Bretthorst L G}{2003}{Probability Theory: The Logic of Science}{Cambridge University Press}{Cambridge}

\end{thebibliography}
\end{document}